\documentclass[amsmath,prl,twocolumn,superscriptaddress]{revtex4}
\usepackage{color}
\usepackage{amsfonts}
\usepackage{amssymb}
\usepackage{graphicx}
\usepackage{pstricks}

\begin{document}
\title{Current-induced forces for nonadiabatic molecular dynamics}
\author{Feng Chen}
\affiliation{Department of Physics, University of California San Diego, La Jolla, CA 92093, USA}
\author{Kuniyuki Miwa}
\affiliation{Department of Chemistry \& Biochemistry, University of California San Diego, La Jolla, CA 92093, USA} 
\affiliation{Surface and Interface Science Laboratory, RIKEN, Wako, Saitama, 351-0198, Japan}
\author{Michael Galperin}
\email{migalperin@ucsd.edu}
\affiliation{Department of Chemistry \& Biochemistry, University of California San Diego, La Jolla, CA 92093, USA} 


\begin{abstract}
We present general first principles derivation of expression for current-induced forces.
The expression is applicable in non-equilibrium molecular systems with arbitrary 
intra-molecular interactions and for any electron-nuclei coupling. It provides a controlled consistent
way to account for quantum effects of nuclear motion, accounts for electronic non-Markov character of
the friction tensor, and opens way to treatments beyond strictly adiabatic approximation.
We show connection of the expression with previous studies,
and discuss effective ways to evaluate the friction tensor. 
\end{abstract}

\maketitle

{\em\bf Introduction.}
Nonadiabatic molecular dynamics (NAMD) is a fundamental problem related to breakdown of the
usual timescale separation (the Born-Oppenheimer approximation) between electron and nuclear
dynamics. The NAMD plays an important role in many processes, ranging from 
chemistry~\cite{GuoJPCL16,SubotnikJPCC15} and photochemistry~\cite{LeiZhuJPhotoChemPhotoBiolA16} 
to spectroscopy~\cite{MukamelJCTC16,SubotnikJCTC15,SubotnikJCP14,PetitSubotnikJCP14} 
and nonradiative electronic relaxation~\cite{LongJPCL16}, and from electron and proton 
transfer~\cite{PrezhdoBeljonnePCCP15,ZhuSciRep16,GanglongThielJCTC16} 
to coherent control~\cite{HenricksenJCP16} and photo-induced energy 
transfer~\cite{CokerAnnRevPhysChem16,TullyWodtkeShaferJPCC15}.
Significance of the problem stems from both complexity of it fundamental theoretical description,
and applicational importance for development of optoelectronic~\cite{OhmuraAIPAdv16,TretiakAccChemRes14} 
and optomechanical~\cite{FilatovJPCL16} molecular devices.

A crucial part of formulating nuclear dynamics is definition of nuclear forces induced
by electronic subsystem. A number of recent studies has discussed ways to account
for ``electronic friction'' in the dynamics~\cite{vonOppenPRL11,vonOppenBJN12,ShenviTullyFaradDisc12,SubotnikJPCB14,NitzanSubotnikJCP15,TullyPRL16,TullyPRB16,SubotnikJCP16,SubotnikPRL17,SubotnikPRB17}. 
Majority of the studies employ intuitive reasoning in the formulations~\cite{ShenviTullyFaradDisc12,SubotnikJPCB14,NitzanSubotnikJCP15,TullyPRL16,TullyPRB16,SubotnikJCP16,SubotnikPRL17,SubotnikPRB17}.
Also, some of the works are confined only to non-interacting and/or equilibrium electronic systems~\cite{ShenviTullyFaradDisc12,SubotnikJPCB14,NitzanSubotnikJCP15,TullyPRL16,TullyPRB16,SubotnikJCP16,SubotnikPRB17}.
A consistent derivation of nuclear forces was presented within path integral formulation employing
the Feynman-Vernon  influence functional~\cite{NewnsPRB95,TodorovPRB12,HedegardBrandbygePRL15}.
The derivation lead to Langevin equation driven by a set of forces:
friction, non-conservative, renormalization, and Berry phase. However, the studies are restricted
to non-interacting electronic systems and 
to linear coupling
between nuclear and electronic degrees of freedom~\cite{TodorovPRB12,HedegardBrandbygePRL15}.  
Finally, all the aforementioned works consider only extremely slow (Ehrenfest) nuclear dynamics.

Here we discuss a derivation of nuclear forces for nonadiabatic nuclear dynamics,
which accounts for nonequilibrium character of a molecular system, intra-molecular interactions,
and general coupling between electronic and nuclear degrees of freedom.
Moreover, the consideration goes beyond extremely slow (Ehrenfest) limit of nuclear motion;
so that resulting expressions are valid also in the intermediate regime and are applicable in
surface hopping considerations. Structure of the paper is as follows: 
first we
introduce a model and present the general derivation. 
After that, we show relation to previous studies indicating corresponding approximations,
establish connection with the Zubarev's method of nonequilibrium statistical operator, and
discuss efficient ways to simulate the forces in nonequilibrium interacting molecular systems. 
Finally, we summarize our findings and indicate directions of further research.

{\em\bf General derivation.}
We consider dynamics of a molecule $M$ adsorbed on a surface(s) $K$.
Hamiltonian of the system is separated into nuclear kinetic and potential energies, 
and electron Hamiltonian which depends on nuclear coordinates $\hat q$
\begin{equation}
\hat H = \sum_{\alpha} \frac{\hat p_\alpha^2}{2} + \hat V(\hat q) + \hat H_e(\hat q)
\end{equation}
$\hat H_e(\hat q)$ consists of molecular $\hat H_M(\hat q)$ and contacts $\hat H_K$ parts, 
and coupling $\hat V_K(\hat q)$ between them.
$\hat H_M(\hat q)$ can include any intra-molecular interactions 
and is represented in the basis of many-body electronic states 
$\{\lvert S\rangle\}$. 
Explicit expressions are
\begin{equation}
\label{He}
\begin{aligned}
&\hat H_M = \sum_{S_1,S_2} \lvert S_1\rangle H^{e}_{S_1S_2}(\hat q) \langle S_2\rvert;\quad 
\hat H_K = \sum_{k\in K} \varepsilon_k\, \hat c_k^\dagger \hat c_k; \\
&\hat V_K = \sum_{k\in K}\sum_{(S_1,S_2)}\bigg( V_{k\,(S_1,S_2)}(\hat q)\, \hat c_k^\dagger\, 
\lvert S_1\rangle\langle S_2\rvert + H.c. \bigg)
\end{aligned}
\end{equation}
Here $\hat c_k^\dagger$ ($\hat c_k$) creates (annihilates) electron in state $k$ of contact $K$,
$S_1$ and $S_2$ indicate a pair of molecular many-body states
which belong to the same charging block ($N_{S_1}=N_{S_2}$) in $\hat H_M$
and to charging blocks different by one electron ($N_{S_1}+1=N_{S_2}$)
in $\hat V_K$ ($N_S$ is number of electrons in state $\lvert S\rangle$).
Note that although $\hat H_M$ is written in the diabatic basis, derivation below is 
not restricted to this particular choice.

Effective evolution for the nuclear density matrix is obtained by tracing 
density matrix of the whole system over electron degrees of freedom 
$\hat \rho_{vib}\equiv\mbox{Tr}_e\{\hat \rho\}$.
Assuming initial density matrix being direct product of nuclear and electron density matrices,
$\hat \rho^{(0)}=\hat \rho_{vib}^{(0)}\otimes\hat\rho_{el}^{(0)}$,
nuclear effective evolution can be represented in terms of the Feynman-Vernon functional 
$F$~\cite{FeynmanVernonAnnPhys63} 
\begin{align}
&\rho_{vib}(2) = \int d1\,\mathcal{K}(2;1)\, \rho_{vib}(1)
\\
\label{K}
&\mathcal{K}(2;1) =
 \int_1^2 D(x,y)\, e^{i \big(S_{vib}(x)-S_{vib}(y)\big)}\, F(x,y)
 \\
 &S_{vib}(q) = \int dt\, L(q,\dot q) \equiv \int dt\bigg(\sum_\alpha\frac{\dot q_\alpha^2}{2}-V(q) \bigg)
 \\
 \label{FV}
& F(x,y)=\bigg\langle T_c\, e^{-i\int_c d\tau\, \hat H_e\big(q(\tau)\big)} \bigg\rangle
\equiv e^{{}\, i\, S_{eff}(x,y)}
\end{align}
Here $i\equiv(x,y)$ ($i=1,2$) represent pair of nuclear coordinates on time-ordered, $x$, 
and anti-time-ordered, $y$, branches of the Keldysh contour, $\mathcal{K}(2,1)$ is the nuclear propagation 
kernel from time $t_1$ to $t_2$, $S_{vib}(q)$ ($q=x,y$) is the action of free nuclear evolution,
$T_c$ is the contour ordering operator, and $\langle\ldots\rangle$ is quantum mechanical and statistical average over only electronic degrees 
of freedom.

Now we want to derive expression for the effective action $S_{eff}$, Eq.~(\ref{FV}).  
To do so we separate classical, $Q_\alpha$,  and quantum, $\xi_\alpha$, nuclear dynamics 
by transferring to the Wigner coordinates
\begin{equation}
 Q_\alpha=\frac{x_\alpha+y_\alpha}{2} \qquad \xi_\alpha=x_\alpha-y_\alpha
\end{equation}
and rewrite the Hamiltonian in the form of zero-order classical nuclear evolution 
plus quantum perturbation 
\begin{equation}
 \hat H_e\big( q(\tau)\big) = \hat H_e\big( Q(\tau)\big) +  \hat V_e\big( q(\tau)\big)
\end{equation} 
where $\hat V_e\big( q(\tau)\big)\equiv \hat H_e\big( q(\tau)\big) -  \hat H_e\big( Q(\tau)\big)$.
Following Ref.~\onlinecite{NewnsPRB95} we replace
$\hat V_e\big( q(\tau)\big)$ by $\lambda \hat V_e\big( q(\tau)\big)$,
and employ the linked cluster theorem. This leads to
\begin{equation}
 \label{LC}
 S_{eff}(x,y) = -\int_0^1 d\lambda \int_c d\tau \big\langle \hat V_e^H\big(q(\tau)\big)\big\rangle_c
\end{equation}
Here superscript $H$ shows that the operator is in the Heisenberg picture, 
and subscript $c$ indicates that one has to consider only connected diagrams.

So far consideration is exact. Now we perform expansion of (\ref{LC}) in $\lambda \hat V_e\big(q(\tau)\big)$.
We justify the expansion in `quantumness' of the nuclei
by noting that for purely classical nuclei, $x=y$, the functional (\ref{FV}) is unity~\cite{Kamenev_2011}
and that for relatively slow nuclear motion deviation from classical trajectory is small.
Expanding (\ref{LC}) up to first order in $\lambda \hat V_e\big(q(\tau)\big)$
and evaluating integrals in $\lambda$ yields
\begin{align}
\label{cumulant}
S_{eff}(x,y) \approx& -\bigg\langle\int_c d\tau\,\hat V_e^I\big(q(\tau)\big)\bigg\rangle_c
\\
+& \frac{i}{2} \bigg\langle T_c\,\int_c d\tau_1\int_c d\tau_2\,\hat V_e^I\big(q(\tau_1)\big)\,\hat V_e^I\big(q(\tau_2)\big)\bigg\rangle_c
\nonumber
\end{align}
where superscript $I$ shows that the operator is in the interaction picture,
i.e. its evolution is defined by Hamiltonian $\hat H_e\big(Q(\tau)\big)$.
Here, first row corresponds to first order and second - to second order of the cumulant expansion.
Employing the Langreth rules~\cite{HaugJauho_2008} to project (\ref{cumulant}) onto real time axis
and  expanding $\hat V_e\big(q(\tau)\big)$ up to second order in the quantum coordinates $\xi_\alpha$ 
leads to
\begin{align}
\label{cumulant_rt}
S_{eff}(x,y) &\approx 
-\sum_\alpha\int dt\; \xi_\alpha(t)\,\big\langle \partial_\alpha \hat H_e^I\big(Q(t)\big)\big\rangle_c
\\
&+ \frac{i}{2} \sum_{\alpha,\beta}\int dt_1\int dt_2 \,\xi_\alpha(t_1)\,\Pi_{\alpha\beta}(t_1,t_2)\,\xi_\beta(t_2)
\nonumber
\end{align}
where
\begin{align}
\label{daHI}
& \big\langle \partial_\alpha \hat H_e^I\big(Q(t)\big)\big\rangle_c
\equiv \mbox{Tr}_e\big\{\partial_\alpha \hat H_e\big(Q(t)\big)\, \hat\rho_{el}(t)\big\}_c
 \\
\label{Pi}
&\Pi_{\alpha\beta}(t_1,t_2) =\frac{1}{2}\bigg\langle 
\bigg\{\partial_\alpha\hat H_e^I\big( Q(t_1)\big); \partial_\beta\hat H_e^I\big( Q(t_2) \big) \bigg\}
\bigg\rangle_c
\end{align}
Here  $\{\ldots;\ldots\}$ is anti-commutator and $\hat H_e\big(Q(t)\big)$ 
is the electronic Hamiltonian in the Schr{\" o}dinger picture build for nuclear frame $Q(t)$. 

Using (\ref{cumulant_rt}) in (\ref{K}) yields
\begin{align}
\label{KQ}
& \mathcal{K}(2;1) = \int D(Q,\xi) \exp\bigg[i\xi_2\dot{Q}_2-i\xi_1\dot{Q}_1\bigg]\times
\\ & 
\exp\bigg[
i \sum_{\alpha}\int dt\, \xi_\alpha(t) \bigg( 
L_{\alpha}(t)+i\sum_{\beta}\int dt'\, \Pi_{\alpha\beta}(t,t')\frac{\xi_\beta(t')}{2}\bigg)\bigg]
\nonumber
\end{align}
where
\begin{equation}
 L_{\alpha}(t)=-\ddot{Q}_\alpha(t)
 -\partial_\alpha V\big(Q(t)\big)-\big\langle\partial_\alpha \hat H_e^I\big(Q(t)\big)\big\rangle
\end{equation}
is the Lagrangian.
Finally, employing the Hubbard-Stratonovich transformation and integrating out 
the quantum coordinates leads to 
\begin{equation}
\label{Langevin}
\ddot{Q}_\alpha(t)
=-\partial_\alpha V\big(Q(t)\big)- \big\langle\partial_\alpha \hat H_e^I\big(Q(t)\big)\big\rangle + f_\alpha(t)
\end{equation}
where $f_\alpha(t)$ is stochastic force which satisfies
\begin{equation}
\label{corr}
\big\langle f_\alpha(t_1)\, f_\beta(t_2)\big\rangle = \Pi_{\alpha\beta}(t_1,t_2)
\end{equation}
Eq.~(\ref{Langevin}) is the stochastic Langevin equation for classical nuclear dynamics driven by
quantum electronic bath. The latter is characterized by Eqs.~(\ref{daHI}), (\ref{Pi}), and (\ref{corr}). 

{\em\bf Friction tensor.}
Eq.~(\ref{Langevin}) is starting point of current-induced  nuclear forces consideration 
in Ref.~\onlinecite{vonOppenBJN12}. 
The consideration was restricted to non-interacting electron systems.

To compare with other results presented in the literature recently, we utilize  
assumption of fast electron dynamics on the timescale of nuclear motion, which allows to transfer
to the reduced description of the nonequilibrium electronic system. In particular, employing Zubarev's 
method of the nonequilibrium statistical operator,
electronic density operator $\hat\rho_{el}(t)$ is expressed in terms of the relevant distribution 
$\hat\rho_{rel}(t)$ as~\cite{Zubarev_1996}
\begin{align}
\label{rho_rel}
 \hat\rho_{el}(t) =& \hat\rho_{rel}(t) 
 - \int_{-\infty}^t dt'\, e^{-\delta(t-t')}\hat U(t,t')
 \\ \times&
 \big\{\partial_{t'}\hat\rho_{rel}(t')
 +i\big[\hat H_{e}\big(Q(t')\big);\hat\rho_{rel}(t')\big]\big\}\hat U^\dagger(t,t')
 \nonumber
\end{align}
where 
$\hat U(t_1,t_2)= T \exp\bigg[-i\int_{t_1}^{t_2} dt\, \hat H_e\big(Q(t)\big)\bigg]$ 
is the electronic evolution operator under classical nuclear driving, $T$ is time-ordering operator
and $\delta\to +0$.

In the limit of extremely slow nuclear driving, when the relevant distribution becomes identical
with steady-state electron distribution, $\hat\rho_{rel}(t')\approx\hat\rho_{ss}^{t'}$
($\rho_{ss}^{t'}$ is steady-state electronic distribution for nuclear frame 
$Q(t')$)~\cite{MorozovRopkeCMP98}, 
so that
 $\big[\hat H_{e}\big(Q(t')\big);\hat\rho_{rel}(t')\big] \approx 0$ and
 $\partial_{t'}\hat\rho_{rel}(t') \approx \sum_\beta \partial_\beta\hat\rho_{ss}^{t'}\, \dot{Q}_\beta(t')$,
using (\ref{rho_rel}) in (\ref{daHI}) yields the
Langevin equation (\ref{Langevin}) in the form
\begin{align}
\label{Langevin_ss}
\ddot{Q}_\alpha(t)
=& -\partial_\alpha V_{eff}\big(Q(t)\big)
\\ &
-\sum_\beta\int dt'\, \gamma_{\alpha\beta}(t,t')\,\dot{Q}_\beta(t') + f_\alpha(t)
\nonumber
\end{align}
Here
\begin{align}
& \partial_\alpha V_{eff}\big(Q(t)\big) \equiv \partial_\alpha V\big(Q(t)\big) +  \mbox{Tr}_e\big\{ \partial_\alpha \hat H_e\big(Q(t)\big)\,\hat\rho_{ss}^t\big\}
\\
\label{gamma}
& \gamma_{\alpha\beta}(t_1,t_2) \equiv -\theta(t_1-t_2)\, e^{-\delta(t_1-t_2)}
\\ &\qquad\times
\mbox{Tr}_e\big\{\partial_\alpha\hat H_e\big(Q(t_1)\big)\,\hat U(t_1,t_2)\,\partial_\beta\hat \rho_{ss}^{t_2}\,\hat U^\dagger(t_1,t_2)\big\}_c
\nonumber
\end{align}
are the effective (renormalized) nuclear force and the friction tensor, respectively.
Note that Markov version of equation (\ref{Langevin_ss})  
was considered in Refs.~\onlinecite{SubotnikPRL17,SubotnikPRB17}.

{\em\bf Fluctuation-dissipation theorem.}
At equilibrium,  when nuclei do not move at the timescale of electron correlation,
$\hat\rho_{ss}\to\hat\rho_{eq}=e^{-\beta\hat H_e(Q)}/Z$,
where $\beta=1/k_BT$ and $Z=\mbox{Tr}_e\big\{e^{-\beta \hat H_e(Q)}\big\}$. 
Employing the Sneddon's formula~\cite{Puri_2001},
Eq.(\ref{gamma}) becomes
\begin{align}
\label{gamma_eq}
&\gamma_{\alpha\beta}(t_1-t_2)
\\
&=\beta\,\theta(t_1-t_2)\int_0^1 dx\,\big\langle \partial_\alpha\hat H_e^I(t_1)\;\partial_\beta 
\hat H_e^I(t_2+i\beta\hbar x)\big\rangle_c
\nonumber \\ &=
 \beta\,\theta(t_1-t_2)\int_0^1 dx\,\big\langle \partial_\beta\hat H_e^I(t_2-i\beta\hbar x)\; \partial_\alpha 
 \hat H_e^I(t_1)\big\rangle_c
\nonumber
\end{align}
This form of $\gamma_{\alpha\beta}$ satisfies the fluctuation-dissipation theorem~\cite{Zubarev_1996} 
(see SM~\cite{SM} for derivation) 
\begin{equation}
\label{fd}
 \Pi_{\alpha\beta}(E) = E\, \mbox{Re}\,\gamma_{\alpha\beta}(E) \,\coth\frac{\beta E}{2},
\end{equation}
which at high temperatures reduces to its classical version
$
 \Pi_{\alpha\beta}(E) = \frac{2}{\beta}\, \mbox{Re}\,\gamma_{\alpha\beta}(E)
$.
Here $\Pi_{\alpha\beta}(E)$ is the Fourier transform of (\ref{Pi}).

{\em\bf Additive electronic Hamiltonian.}
A number of studies~\cite{NewnsPRB95,vonOppenPRL11,vonOppenBJN12,TodorovPRB12} considered
electronic Hamiltonian (\ref{He}) consisting of  zero-order coordinate-independent part and
electron-nuclear coupling depending on nuclear coordinates
\begin{equation}
\label{Headd}
 \hat H_e(\hat q) = \hat H_e^{(0)} + \hat H_e^{(1)}(\hat q)
\end{equation} 
In this case, we can expand evolution operator $\hat U$ up to linear
order in $\hat H_e^{(1)}$ as 
$\hat U(t_1,t_2) \approx \hat U_0(t_1,t_2) -
i\int_{t_2}^{t_1}dt'\, \hat U_0(t_1,t')\, \hat H_e^{(1)}\big(Q(t')\big)\,
\hat U_0(t',t_2)$,
where
$\hat U_0(t_1,t_2) = \exp\big[-i\,\hat H_e^{(0)}(t_1-t_2)\big]$.
Substituting the expansion into (\ref{daHI}) and keeping only terms up to second order in 
$\hat H_e^{(1)}$ leads to
\begin{align}
\label{daHI_H0}
 &\big\langle \partial_\alpha \hat H_e^{I}\big(Q(t)\big)\big\rangle_c
 \approx \big\langle \partial_\alpha \hat H_e^{(1)I}\big(Q(t)\big)\big\rangle
 + \int dt'\,\Pi^r_\alpha(t,t')
 \\
\label{Pir}
&\Pi^r_\alpha(t,t')\equiv -  i\,\theta(t-t')\big\langle \big[\partial_\alpha \hat H_e^{(1) I}\big(Q(t)\big);
 \hat H_e^{(1) I}\big(Q(t')\big)\big]\big\rangle_{c}
\end{align}
Here superscript $I$ indicates interaction picture with respect to Hamiltonian $\hat H_e^{(0)}$.
As previously, first term on the right in (\ref{daHI_H0}) renormalizes nuclear potential,
while second term accounts for electronic friction.
Utilizing second order expansion in $H_e^{(1)}$ also in (\ref{Pi})
(i.e. using $\hat U_0$ in place of $\hat U$ in the expression) 
generalizes  non-interacting considerations of Refs.~\onlinecite{NewnsPRB95,vonOppenPRL11,vonOppenBJN12,TodorovPRB12} 
to the case of interacting non-equilibrium systems.
As an example in SM~\cite{SM} we use (\ref{Pi}) and (\ref{Pir}) to derive results of  Ref.~\cite{TodorovPRB12}.

Previously introduced in the literature standard friction, nonconservative, renormalization and 
Berry phase forces are related to our electronic friction as follows.
Expressions for the friction, Eqs.~(\ref{daHI}), (\ref{gamma}), or (\ref{Pir}),
have generic form $F(t,t')=\theta(t-t')f(t,t')$.
At steady state, when $f(t,t')=f(t-t')$, Fourier transform of $F(t,t')$ is
\begin{equation}
 F(E) = 
 -i\, PP\int\frac{d\omega}{2\pi}\frac{f(\omega)}{E-\omega} + \pi f(E)
\end{equation}
where $PP$ is the principle part.
The standard friction is identified with $\pi\, \mbox{Im}\, f(E)$,
the nonconservative force is given by $\pi\, \mbox{Re}\, f(E)$,
the renormalization contribution comes from 
$PP\int\frac{d\omega}{2\pi} \frac{\mbox{Im}\, f(\omega)}{E-\omega}$,
and the Berry phase force is associated with
$PP\int\frac{d\omega}{2\pi} \frac{\mbox{Re}\, f(\omega)}{E-\omega}$
(see Ref.~\onlinecite{TodorovPRB12} for details).

\begin{figure}[t]
\centering\includegraphics[width=\linewidth]{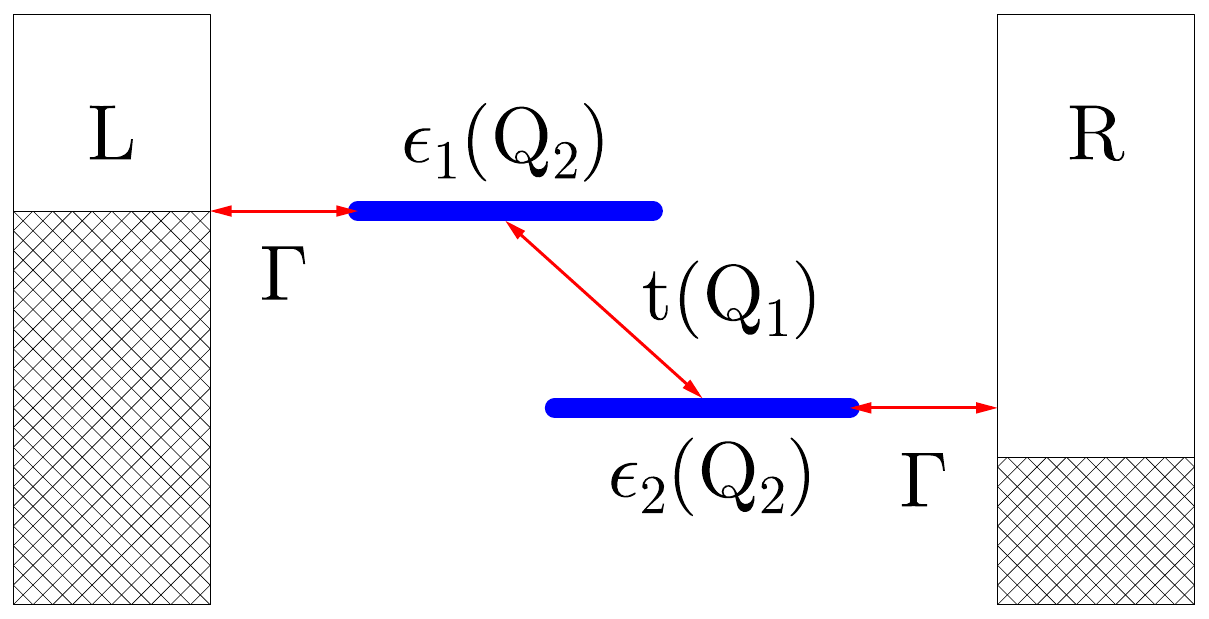}
\caption{\label{fig1m}
Two-sites two-modes non-interacting junction model of Ref.~\cite{TodorovPRB12}.
}
\end{figure}

{\em\bf Effective evaluation of the friction tensor.}
We now discuss effective ways to evaluate current-induced forces in interacting molecular 
junctions. 
As discussed in our previous 
publications~\cite{WhiteOchoaMGJPCC14,MGANJPCL15,MGChemSocRev17},
methods of the non-equilibrium atomic limit (formulated in the basis of many-body states of 
the molecule) may be a convenient alternative to standard (orbital based) treatment of electronic degrees of freedom. In particular, recently introduced by us non-equilibrium diagrammatic technique 
for the Hubbard Green functions~\cite{ChenOchoaMGJCP17,MiwaChenMGSciRep17}
may be beneficial also for evaluation of the friction tensor.

Deferring study of non-adiabatic effects on the timescale of electronic correlations to future research, 
we focus on the additive electronic Hamiltonian (\ref{Headd}) with nuclear coordinate 
dependence confined only to molecular Hamiltonian $\hat H_M$.
In this case the dissipation term, Eq.~(\ref{Pir}), is 
\begin{align}
 \label{Piradd}
 \Pi^r_\alpha(t_1,t_2) =& \sum_{S_1,S_2,S_3,S_4}\partial_\alpha H^e_{S_1S_2}(Q)\; H^e_{S_4S_3}(Q)\,
 \\ &\qquad\qquad\times
 \mathcal{D}^r_{S_1S_2,S_3S_4}(t_1,t_2)
\nonumber  
\end{align}
where $\mathcal{D}^r$ is the retarded projection of the single-particle Hubbard Green function
\begin{equation}
\label{HubGF}
 \mathcal{D}_{S_1S_2,S_3S_4}(\tau_1,\tau_2) = -i\big\langle T_c\, \hat X_{S_1S_2}(\tau_1)\,
 \hat X_{S_3S_4}^\dagger(\tau_2)\big\rangle_e
\end{equation}
Here $\hat X_{S_1S_2}(\tau)$ ($\hat X_{S_1S_2}\equiv\lvert S_1\rangle\langle S_2\rvert$) 
is the Hubbard (projection) operator, $\tau_{1,2}$ are the contour variables.
Green function (\ref{HubGF}) can be evaluated
utilizing non-equilibrium diagrammatic technique of Ref.~\onlinecite{ChenOchoaMGJCP17}.
Note that while friction is expressed in terms of single-particle Hubbard Green function,
similar orbital based treatment inevitably leads to appearance of a two-particle Green function.
Note also that Hubbard GF treats formally exactly all electron correlations within the molecule 
(and accounts approximately correlations between molecule and contacts),
while orbital based consideration of intra-molecular interactions is a complicated numerical task. 

{\em\bf Numerical example.}
As an illustration we consider two-sites two-vibrational modes non-interacting junction 
model of Ref.~\cite{TodorovPRB12} (see Fig.~\ref{fig1m}). 
Hamiltonian (\ref{Headd}) takes the form
$H_e^{(0)}=\sum_{m=1,2}\epsilon_m\hat d_m^\dagger\hat d_m -t(\hat d_1^\dagger\hat d_2+H.c.)+\sum_{k\in L,R}\varepsilon_k\hat c_k^\dagger\hat c_k+\sum_{\ell\int L}(V_{\ell 1}\hat c_\ell^\dagger\hat d_1+H.c.)+\sum_{r\in R}(V_{r2}\hat c_r^\dagger\hat d_2+H.c.)$ and 
$H_e^{(1)}(Q)=m_1(\hat d_1^\dagger\hat d_2+H.c.)Q_1+m_2(\hat d_1^\dagger\hat d_1-\hat d_2^\dagger\hat d_2)Q_2$. Here $\hat d^\dagger_m$ ($\hat d_m$) creates (annihilates) electron in 
orbital $m$ ($m=1,2$).
For this form of the Hamiltonian (\ref{Pir}) yields friction tensor
$\Pi^r_{\alpha\beta}(t,t')=-i\theta(t-t')\big\langle\big[\partial_\alpha \hat H_e^{(1) I}\big(Q\big);
 \partial_\beta\hat H_e^{(1) I}\big(Q\big)\big]\big\rangle_{c}$.
 To make comparison with Ref.~\onlinecite{TodorovPRB12} easier,
we simulate $\Lambda(E)$ function, which is related to the friction tensor as
$
 \Pi^r_{\alpha\beta}(t,t')=\theta(t-t')\, 2\pi i\, \Lambda_{\alpha\beta}(t,t')
$.

We compare NEGF results for current-induced forces (exact for the model) with the
Hubbard NEGF simulations. Green function (\ref{HubGF})
was simulated within second order diagrammatic perturbation theory
(see Ref.~\cite{ChenOchoaMGJCP17} for details).
Parameters of the simulation are $T=300$~K, $\epsilon_1=-\epsilon_2\equiv\epsilon_0=0.1$~eV, 
$t=0.2$~eV, $\Gamma=1$~eV, and  $m_1=m_2=0.01$~eV/AMU${}^{1/2}$\,\AA. 
Fermi energy is taken as origin, $E_F=0$. Simulations are performed 
for bias $V=1$~V;  $\mu_{L,R}=\pm |e|V/2$. 

\begin{figure}[t]
\centering\includegraphics[width=\linewidth]{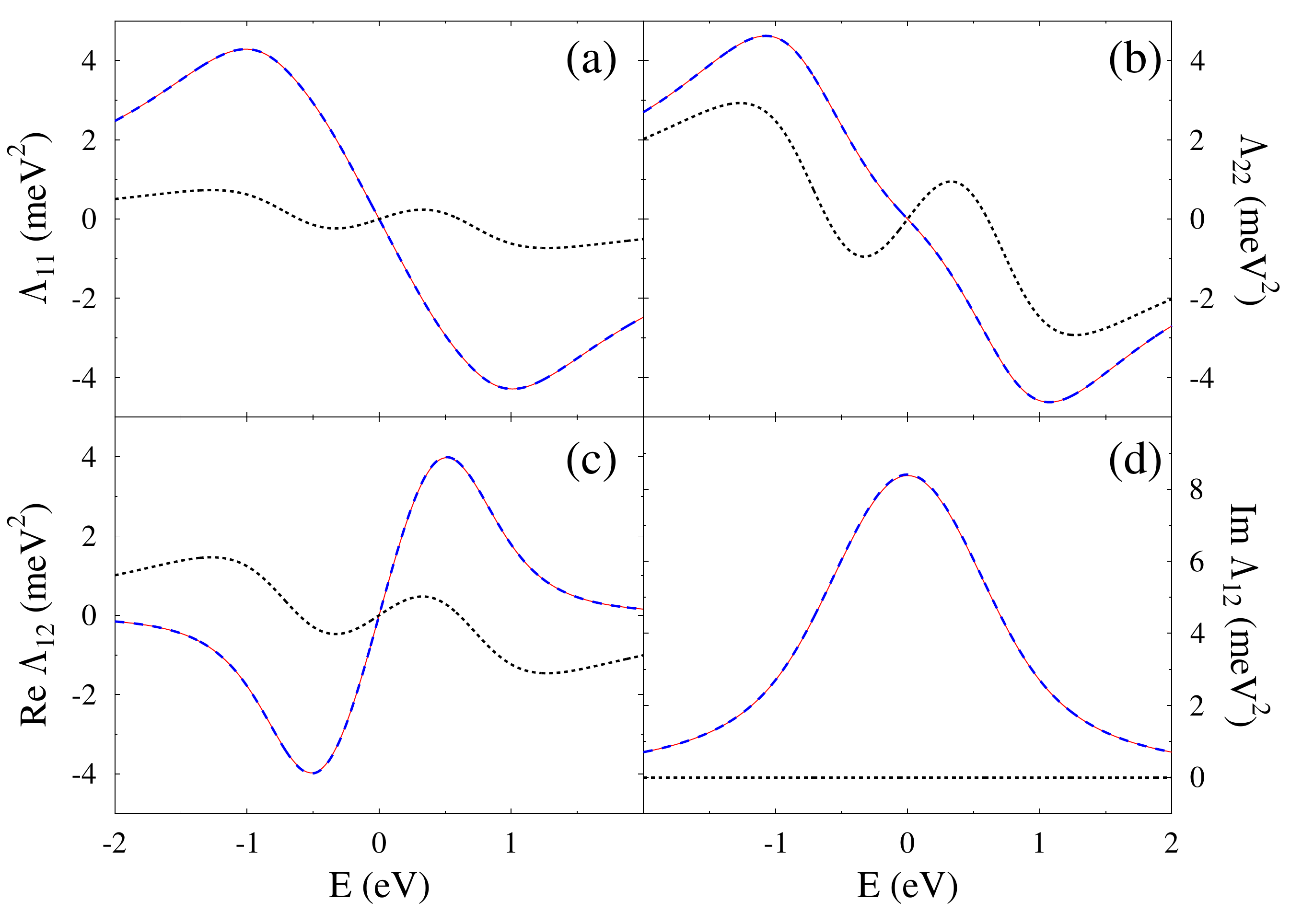}
\caption{\label{fig2m}
Friction tensor $\Lambda(E)$ for the noninteracting junction model 
(Fig.~\ref{fig1m}). Results
of the NEGF calculation~\cite{TodorovPRB12} (solid line, red) are compared with the Hubbard NEGF
(dashed line, blue) and non-equilibrium generalization of the Head-Gordon and Tully
electronic friction~\cite{TullyJCP95} (dotted line, black). See text for parameters.
}
\end{figure}

Figure~\ref{fig2m} shows elements of $\Lambda_{\alpha\beta}(E)$ as obtained
from the NEGF and Hubbard NEGF simulations. It is interesting to note that although
system-bath coupling $\Gamma$ is not small and while the Hubbard NEGF is perturbative 
in system-bath coupling strength expansion, Hubbard simulations follow exact (for the model)
NEGF results very closely. Similar accuracy in a wide range of parameters was noted in our 
recent publication~\onlinecite{MiwaChenMGSciRep17}. We attributed the effect to similarity of
diagrammatic techniques for the two Green functions. 

We also compare the results with 
a generalized version of the Head-Gordon and Tully friction tensor~\cite{TullyJCP95}.
In our model this generalization is given by $D^r_{ab,ba}$ and $D^r_{ba,ba}$ contributions
in Eq.~(\ref{Piradd}). Here $a$ and $b$ are molecular many-body states with one electron
being in one of eigenstates of the Hamiltonian $H^0$. 
As expected, at non-negligible system-bath coupling this form of the tensor
fails to reproduce correct behavior (see Fig.~\ref{fig2m}). 
Naturally, with decrease in the system-bath coupling $\Gamma$
the Head-Gordon and Tully result becomes more accurate (especially at the energies corresponding
to transitions between molecular eigenstates; see Fig.~S1 in SM~\cite{SM}).

\begin{figure}[t]
\centering\includegraphics[width=0.95\linewidth]{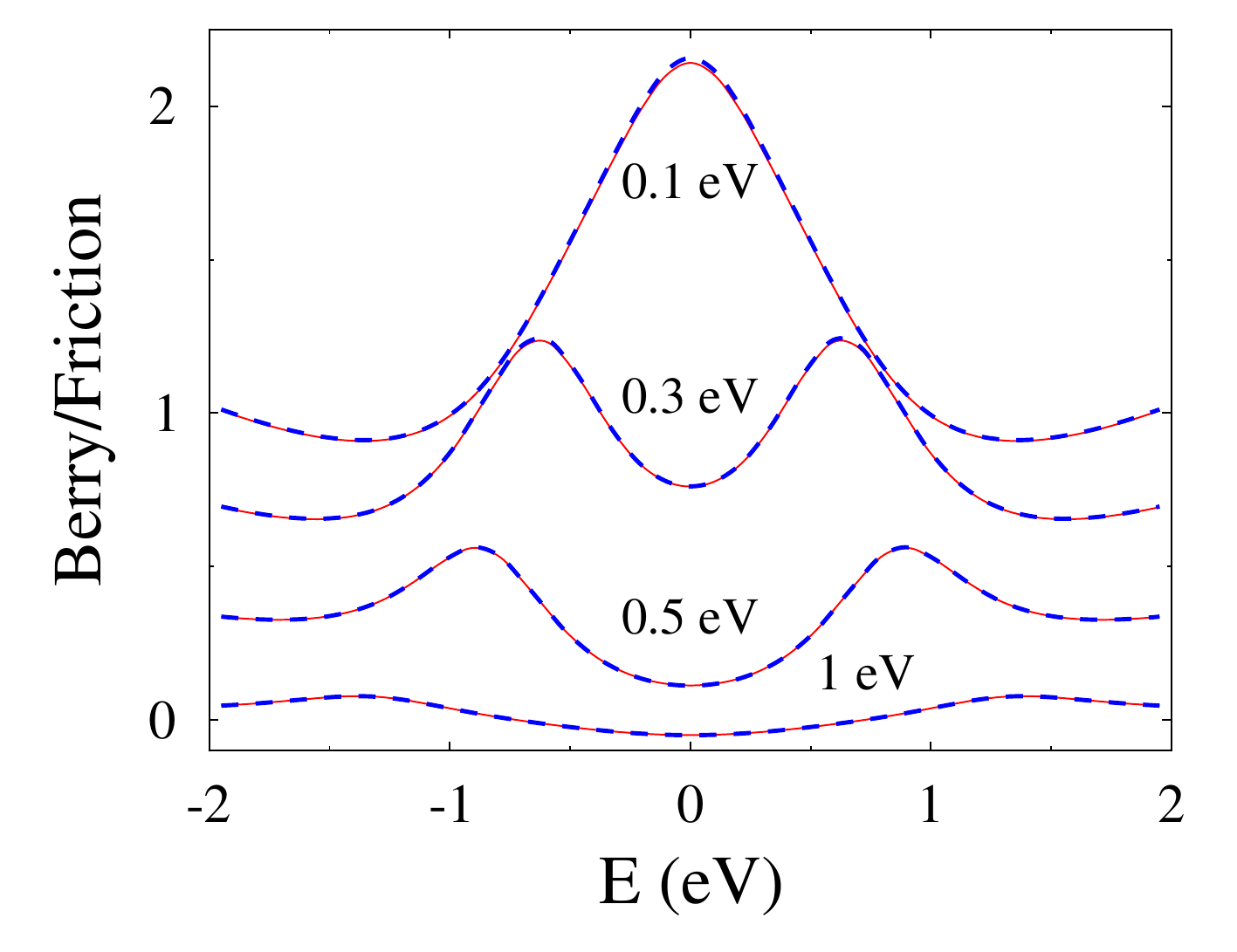}
\caption{\label{fig3m}
The Berry phase force relative to the average friction.
Results of the NEGF calculation~\cite{TodorovPRB12} (solid line, red) are compared with the Hubbard NEGF
(dashed line, blue).
Numbers indicate position of the level $\epsilon_0$.
Parameters are as in Fig.~\ref{fig2m}.
}
\end{figure}

Figure~\ref{fig3m} demonstrates relative importance of the Berry phase force. 
Also here, the Hubbard NEGF calculations closely follow exact (for the model) NEGF results.
As was first indicated in Ref.~\onlinecite{TodorovPRB12}, Berry force is pronounced at energies 
corresponding to transitions between molecular many-body states.
Its significance decreases with separation between transition energy and chemical potential. 

Finally, short discussion of non-Condon effects in current-induced forces is given in SM~\cite{SM}. 
We postpone detailed study to future publication. 

{\em\bf Conclusion.}
We presented general derivation of current-induced forces for non-adiabatic nuclear dynamics
and compared it to previous works. Our derivation goes beyond the usually assumed
extremely slow (Ehrenfest) nuclear dynamics. Thus, resulting expressions for the forces
are applicable also in intermediate regime (for example, in combination with surface-hopping schemes).
The derivation is completely general in a sense that it is applicable in equilibrium and 
non-equilibrium molecular systems which may be open or closed, with or without intra-molecular
interactions (e.g., electron-electron repulsion) taken into account, and for any electron-nuclear coupling.
The derivation is based on a standard cumulant expansion, i.e. it is a first principles consideration,
which allows for consistent extension of the treatment into higher (than the utilized second) orders.
Results of previous works follow from our derivation as particular limiting cases.
We established connection with the Zubarev's method of nonlinear statistical operator. 
This opens a practical way for considerations beyond strictly adiabatic limit.
We also discussed effective ways of evaluating friction tensor in interacting nonequilibrium systems.
In particular, we show that recently introduced by us nonequilibrium diagrammatic technique for 
the Hubbard Green functions~\cite{ChenOchoaMGJCP17} may be a convenient tool 
for evaluation of the friction tensor. For example, usual expressions for the tensor 
in interacting systems require consideration of two-particle Green function; the same consideration
requires evaluation of only single-particle Hubbard Green function. 
Main goal of this study is to demonstrate a consistent general derivation. 
Application of the methodology to actual calculations and elucidation of the role of non-adiabatic 
driving are goals of future research. 


\begin{acknowledgments}
We thank Abraham Nitzan for helpful discussions.
This material is based upon work supported by the National Science Foundation 
under CHE-1565939 and by the Department of Energy under DE-SC0018201.
\end{acknowledgments}



\end{document}